# Pressure Tuned 2D Superconductivity in Black Phosphorus

Meiling Jin[1,5#], Qing Wang[2,3#], Ying Liu[1], Qunfei Zheng[1], Chenkai Li[1], Shaoheng Wang[1], Shanmin Wang[1], Ning Hao[2*], Yuki Nakamoto[4], Katsuya Shimizu[4*], Jinlong Zhu[1*]

[1]*Department of Physics, Southern University of Science and Technology, Shenzhen 518055, China*

[2]*Anhui Key Laboratory of Condensed Matter Physics at Extreme Conditions, High Magnetic Field Laboratory, HFIPS, Anhui, Chinese Academy of Sciences, Hefei, 230031, China*

[3]*Science Island Branch of Graduate School, University of Science and Technology of China, Hefei, Anhui 230026, China*

[4]*Center for Science and Technology under Extreme Conditions, Graduate School of Engineering Science, Osaka University, 1-3 Machikaneyama-Cho, Toyonaka, Osaka, 560-8531, Japan*

[5]*Center for Quantum Physics, School of Physics, Beijing Institute of Technology, Beijing, 100081, China*

Equal contribution: Meiling Jin and Qing Wang

Corresponding authors: Katsuya Shimizu, shimizu.katsuya.es@osaka-u.ac.jp

Ning Hao, haon@hmfl.ac.cn;

Jinlong Zhu, zhujl@sustech.edu.cn

## Abstract

This paper examines the micro-parameters of superconductors. It studies the modulations from weak van der Waals interaction to strong covalence bonding of superconductors. In particular, we studied layered black phosphorus (BP) as a function of pressure. These results reveal a rich scenario of phase transitions and related quantum phenomena, which show that the phases exhibit superconducting states at a pressure higher than 5.0 GPa. In addition, they indicate an angle-dependent upper critical field that demonstrates the dimensional characteristics of superconductivities. This result suggests that the A17 and cubic phases are three-dimensional (3D). The A7 phase shows a two-dimensional (2D) character. The 2D behavior is related to a weakened, distorted, entangled interlayer coupling.

## 1. Introduction

Two-dimensional (2D) superconductivity can be achieved through spatial confinement in thin film forms, 2D electron gas at interfaces, and exfoliation. Early experimental work on amorphous Bi films demonstrated 2D superconductivity through precise control of the superconducting layer thickness (1). Further developments in thin-film growth have led to the observation of high-quality 2D superconductors formed at interfaces between (111)-oriented $KTaO_3$ and the insulating overlayers (2). Recently, developments in the exfoliation of van der Waals (vdW) layered materials have made atomically thin 2D superconductors readily accessible (e.g., $TaS_2$ and $NbSe_2$) (3, 4). It is controversial and rare to achieve 2D superconducting in bulk crystals. Although having 2D characterization, most bulk-layered compounds do not possess these unique properties. The recent upsurge of layered materials brings this topic to the frontier. Quasi-2D superconductivity has been observed (5) in bulk $AuTe_2Se_{4/3}$ pressure-induced crossover from two- to three-dimensions (2D to 3D). Superconducting states have been observed in optimally doped $Bi_2Sr_2CaCu_2O_{8+\delta}$ bulk superconductors (6). Clean-limit 2D superconductivity has been achieved in a bulk single-crystal superlattice formed with a commensurate block layer (7).

Phosphorus shows many polymorphic forms by changing temperature, pressure, or both. Among them, black phosphorus (BP), an elemental narrow-gap semiconductor with a direct band gap (8), received much attention recently as a representative 2D material together with graphene and transition-metal dichalcogenides (TMD) (9-14). BP can be converted from white phosphorus at 12 Kbar and 200 °C (15) and red phosphorus at ~85 Kbar (16). Its sizeable single crystal can be obtained at high pressure (HP), high temperature (HT) (17-19), and by chemical reaction (20). The recent interest in BP is due to its layered structure (A17) and its rich properties in electrocatalysis (21). This structure appears in pressure-driven superconductivity (22-26) and topological phase transitions (27-29). Pressure can be critical in adjusting its structures and novel properties for elementary substances, such as BP. It can provide a way to find quantum fluctuations or quantum critical points.

The phase transition of BP at room temperature (RT) and at 21 K is well established: at room temperature, a layered A7 phase stabilized at 5 to 10.0 GPa, a cubic phase at a pressure higher than 10.0 GPa, and a topological transition at ~1.0 GPa (22-24,30); at 21 K, the A7 phase appears at ~10.0 GPa, and the cubic phase emerges at a pressure higher than 15.0 GPa (31). From the viewpoint of crystal structure, the A17 phase is a Van de Waals stacked layered crystal among buckled individual layers. Each shows Armchair and Zig-Zag directions along *c* and *a*, respectively. At the same time, the A7 phase emerges by applying external pressure and hosts a Zig-Zag configuration of every single layer. As pressure increases above 25.0 GPa, the cubic phase shows a phase coexisting with the A7 phase over an extensive range. It is critically important to determine the accurate pressure boundary at a temperature near the superconducting phase to identify the hosting parent crystal structure. The layered A17 and A7 phases can disclose fundamental mechanisms, such as the connection between anisotropy's strength and the dimensions of the crystal.

## 2. Materials and methods

2.1 Growth of black phosphorus

The black phosphorus was prepared by amorphous red phosphorous under high pressure and high-temperature condition. The starting material, red phosphorus (Aladdin Industrial Corporation, 99.999%), was compacted into pellets, stacked with an hBN capsule, and assembled with the prepared cell parts.[1] Then the cell was placed into a DS × 10 MN cubic press, gradually compressed to 2.0 GPa, and heated to 1000 °C with a holding time of 30 minutes, followed by slowly cooling to room temperature within 10 hours.

2.2 High-pressure measurements

The pressure was generated by a diamond anvil cell (DAC) that consists of two opposing anvils sitting on the supporting plates. Diamond anvils with culet diameters of 400 μm were used for this study. The T301 stainless steel gasket was pre-indented from the thickness of 250 to 40 μm and then drilled with a hole of 150 μm in diameter. The four-probe method was applied for all high-pressure transport measurements. The cubic BN used as the insulating layer was pressed into this hole and further drilled a small center hole with a diameter of 100 μm to serve as a sample chamber, in which NaCl fine powders were used as pressure transmitting medium. High-pressure XRD experiments were performed at SPring-8 beamlines. A monochromatic X-ray beam with a wavelength of 0.4129 Å was chosen for XRD measurements. The pressure was determined by the ruby fluorescence method.

2.3 First-principles calculation

The first-principle calculation was performed using the QUANTUM ESPRESSO package (32) with the projector-augmented wave method (33). In all analyses, the exchange-correlation potential was treated within a generalized gradient approximation (GGA) with Perdew-Burke-Ernzerhof (PBE) (34) functionals. The Brillouin-zone (BZ) sampling was created by using an 11×11×5 Monkhorst-Pack k-point mesh for atomic geometry optimization and self-consistent calculations and a 45×45×15 Monkhorst-Pack k-point mesh for three-dimensional Fermi surface calculations.

## 3. Results and discussion

In this manuscript, a high-quality BP was obtained by HP/HT synthesis. The angle-dependent transport properties up to 44.0 GPa were measured. The phase boundary under high pressure at temperature ~10 K was determined by synchrotron X-ray diffraction with pressures up to 40.0 GPa. The A17 to A7 transition was found at ~10.0 GPa and A7 to cubic phase at pressure ranges of 25 to 40.0 GPa [see Supplementary Fig. S1]. It has been determined that all three phases of BP show superconductive behaviors at $T_c$ lower than 10 K. The dimensional evolution of superconductivity in all three phases has been discussed for the first time. The I-V relationship at low temperatures shows the BKT transition in the A7 phase, demonstrating the 2D electronic structure characterization.

The research on anisotropic superconductivity was first conducted by measuring the upper critical fields parallel ($\mu_0 H_{c\|}$) and perpendicular ($\mu_0 H_{c\perp}$) to the sample plane [see Fig. 1(a-c)], namely in-plane and out of plane, and well described by the Ginzburg Landau (G-L) theory,

$$\mu_0 H_{c\|}(T) = \frac{\Phi_0}{2\pi \xi^{in}(0)\xi^{out}(0)}[1 - \frac{T^2}{T_c^2}] \quad (1)$$

$$\mu_0 H_{c\perp}(T) = \frac{\Phi_0}{2\pi (\xi^{in})^2(0)}[1 - \frac{T^2}{T_c^2}] \quad (2)$$

where $\Phi_0$ is the magnetic flux quantum and $\xi^{in}(0)$ and $\xi^{out}(0)$ are the G-L coherence length at $T$= 0 K. The fits to $\mu_0 H_{c\|}$ and $\mu_0 H_{c\perp}$ of different pressures give anisotropy ratio $\gamma_H = \mu_0 H_{c\|}/\mu_0 H_{c\perp}$ [see Fig. 1(e)]. The upper critical fields of 10.0-23.5 GPa are the highest and distinctly anisotropic with $\gamma_H$ up to 4.5 (comparable with quasi-2D cuprate superconductors $YBa_2Cu_3O_{7-\sigma}$, anisotropy ratio is 3-7; quasi-2D superconductor $NbSe_2$, $MgB_2$, anisotropy ratio is 3.3 and 4.5), while $\gamma_H$ ranges from 1 to 2 in other pressures. The coherence length $\xi^{in}(0)$ and $\xi^{out}(0)$ are between 10 nm and 60 nm in the range of measured pressures, while the shortest coherence length emerged at 10 to 23.5 GPa [see Fig. S2].

Angle-dependent upper critical field $\mu_0 H_c$ measurements were also carried out to confirm the dimensional nature of the observed superconducting state. Interestingly, $\mu_0 H_c$ at 1.6 K is highly angle $\theta$-dependent and varies dome-shaped in the pressure range of 10 to 23.5 GPa [see Fig. 1(f)]. The measured $\mu_0 H_c$ versus $\theta$ are plotted in Fig. 1(g-i), and insets show resistance $R$ versus magnetic field $H$ at different angle $\theta$. We fitted the $\mu_0 H_c$-$\theta$ curves using both the 2D Tinkham model (Equation. (3)), and 3D Ginzburg-Landau anisotropic model (Equation. (4)),

$$(\frac{H_c(\theta)\cos(\theta)}{H_{c\|}})^2 + |\frac{H_c(\theta)\sin(\theta)}{H_{c\perp}}| = 1 \quad (3)$$

$$(\frac{H_c(\theta)\cos(\theta)}{H_{c\|}})^2 + (\frac{H_c(\theta)\sin(\theta)}{H_{c\perp}})^2 = 1 \quad (4)$$

where $\mu_0 H_{c\|}$ and $\mu_0 H_{c\perp}$ are the upper critical fields at $\theta$ = 0° and 90°, respectively. Fig. 1(g, i) shows that the fit agrees well with the 3D Ginzburg-Landau anisotropic model. It demonstrates the 3D nature of superconductivity. Fig. 1(h) shows that the fitting curve agrees well with the 2D Tinkham model, indicating the 2D nature of superconductivity. The above indicated that the superconductivity of BP under pressure underwent an evolution process from 3D to 2D and then to 3D.

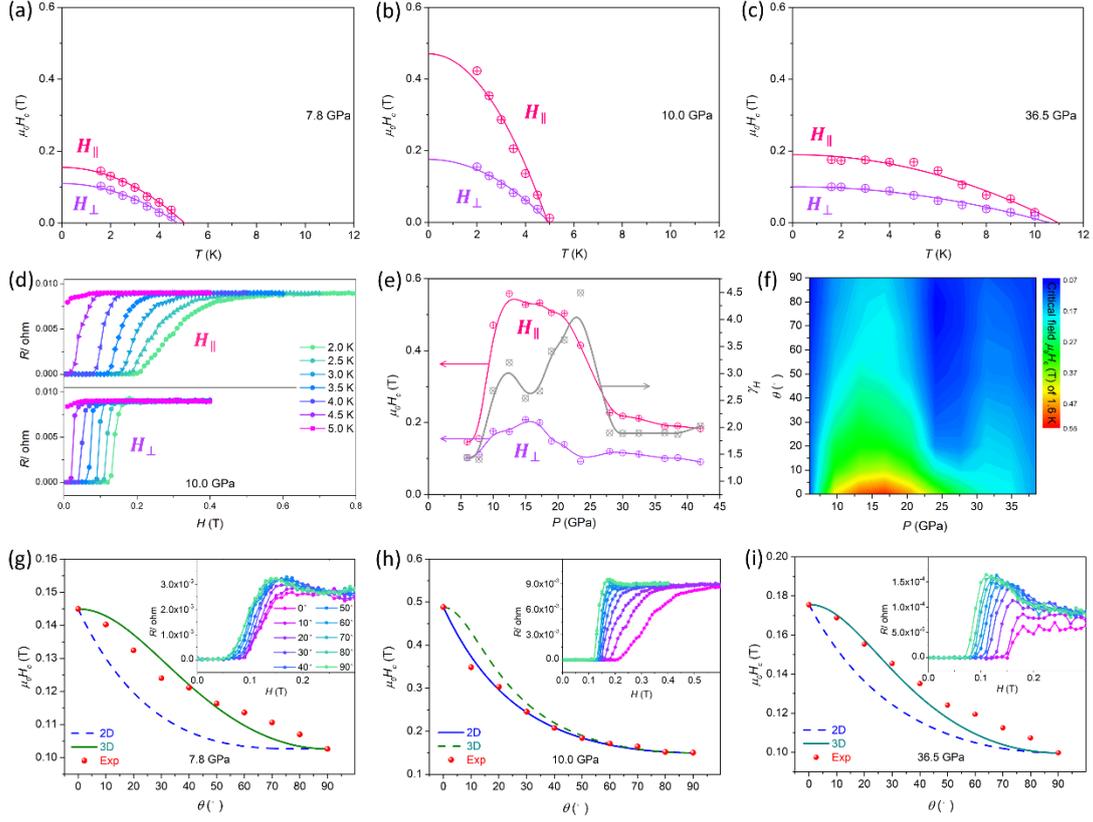

**Fig. 1. The anisotropic superconductivity of a single crystal BP under pressure.** The temperature dependence of these fields for the following in-plane and out-of-plane measurements: (**a**) 7.8 GPa, (**b**) 10.0 GPa, and (**c**) 36.5 GPa. These measurements were made at 90% of $R_N$ (the normal state resistance). (**d**) The resistance at a pressure of 10.0 GPa was measured at different temperatures as a function of the magnetic fields in two directions. (**e**) Pressure dependence of the upper critical field $\mu_0 H_{c\parallel}$, $\mu_0 H_{c\perp}$ and anisotropy is given by the factor $\gamma_H = \mu_0 H_{c\parallel}/\mu_0 H_{c\perp}$. (**f**) Pressure versus angle-dependent upper critical field at 1.6 K. (**g, i**) The blue dashed line, and the solid green line fit the 2D Tinkham model and 3D Ginzburg-Landau anisotropic model, respectively. The $\mu_0 H_c$ curve agrees well with the 3D Ginzburg-Landau anisotropic model for the A17 phase (**g**) and the cubic phase (**i**). (**h**) The solid blue and green dashed lines fit the 2D Tinkham model and 3D Ginzburg-Landau anisotropic model, respectively. The $\mu_0 H_c$ curve agrees well with the 2D Tinkham model for the A7 phase.

The measured current-voltage (I-V) characteristics at different pressures and temperatures are shown in Fig. 2(a, b, c), and the corresponding ln-ln scales were plotted in Fig. 2(d, e, f). The critical currents decrease with increasing temperature. As the temperature raised to near $T_c$, we observed a gradual onset of a resistive state at low currents. The slope of the I-V characteristics evolves smoothly from the normal ohmic state, $V \propto I$, toward a steeper power law, $V \propto I^\alpha (\alpha > 1)$, as superconductivity sets at lower temperatures. At 8.0 GPa, the power law exponent α is less than 3, and the lower the temperature, the lower the rate of change. However, at a pressure of 10 GPa, α passes through the value of three, and the temperature change rate is much larger. This evolution may be interpreted as a Berezinskii–Kosterlitz–Thouless (BKT) transition in a 2D superconductor (35). The onset of a nonlinear I-V characteristic in the original superconducting state is caused by the current-driven unbinding of vortex-antivortex pairs. Thermal fluctuations create these pairs at finite temperatures. This behavior causes the exponent α to smoothly increase as the temperature is lowered from one near the onset of superconductivity to three at $T_{BKT}$ (36, 37). At pressures above 38.0 GPa, the maximum of α drops to three. When α is close to three, it changes slowly with temperature. This behavior suggests a weak relationship between superconductivity and the BKT phase transition. The same experiments were conducted at 7 GPa, 20.4 GPa, and 44.0 GPa (see Fig. S3). The results are consistent with the data obtained in Fig. 2.

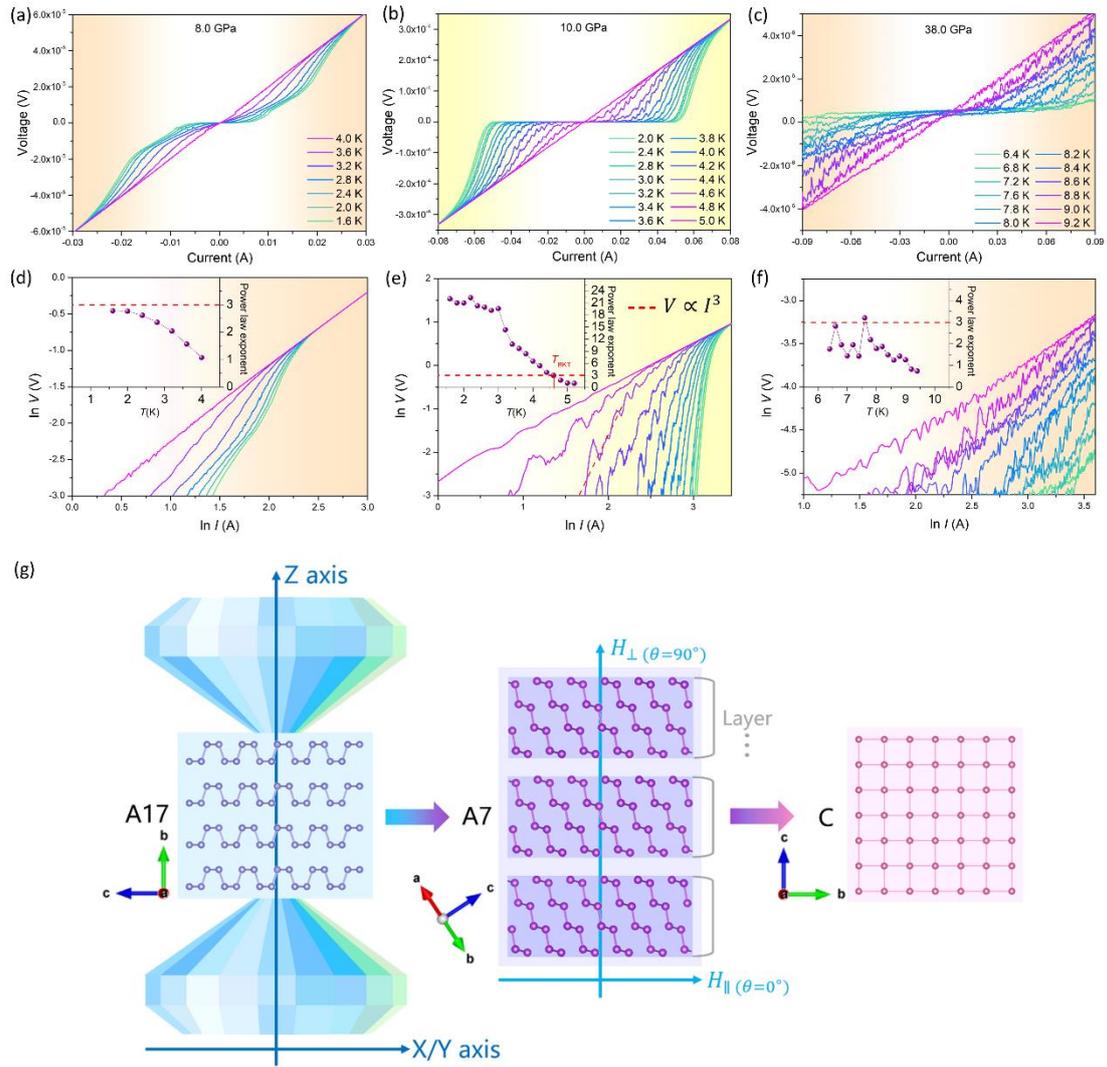

**Fig. 2| BKT transition.** Voltage-current (V-I) characteristics measured at different pressures, 8.0 GPa (**a, d**), 10.0 GPa (**b, e**), 38.0 GPa (**c, f**). The red dashed line in the ln-ln scale corresponds to an $I^3$ dependence. This shape indicates that the Berezinskii-Kosterlitz-Thouless (BKT) transitions are 10.0 GPa (**e**). (**g**) Schematic diagram of crystal structure evolution under pressure. The sample plane of A17 phase is parallel with the diamond culets. In A7 phase, the surface of a nanoflake is also parallel with the diamond culets. The $H_\parallel$ and $H_\perp$ as mentioned above are parallel and perpendicular

with the surface of the nanoflakes, respectively.

To further understand the transport properties and verify the 2D SC in BP, we have carefully performed an *ab* initio calculation to obtain the electronic band structure. The band structure is calculated by density functional theory. The analysis of the standard layered structure for the A7 phase at 8.0 GPa shows that bulk BP has a trivial three-dimensional (3D) Fermi surface [see Fig. S4(a-c)]. In addition, if a standard A7 bulk structure is formed, the angle between the van der Waals layer and the diamond culets will be about 56 degrees, and there will be no minimum or maximum resistance when the magnetic field is perpendicular or parallel with the sample surface. So we infer that there is no standard bulk structure in A7 phase. As the pressure may induce a slip between the layers of pristine A17 to form a nanoflake, we deduced that in A7 phase there may exist nanoflakes whose surfaces are parallel with diamond culets, and there may be nanoflakes with various thickness layers if they're less than the coherent length. For example, two monolayer films from A17 phase can experience being coupled, structural phase transition and forming nanoflakes (see Fig.2g).

Since bulk superconductor exists in bulk crystal consisting of many layers (thickness of 5 μm here for BP, ~$10^4$ layers) and is not a two-dimensional system spatially, a comprehensive theory must treat the interlayer coupling in some appropriate way. In a theoretical study of layered spin systems, which did not treat the magnetic coupling between layers, Hikami and Tsuneto (38) showed that an independent vortex pair configuration in each layer can be energetically favorable relative to a multilayer vortex ring configuration. This situation occurs when the vortex pair separation is less than a critical distance

$$r_0 = \xi(K/K_c)^{1/2} \quad (5)$$

$\xi$ is the Ginzburg-Landau coherence length, which would be the in-plane coherence length in this instance, $K$ is the renormalized stiffness constant ($\pi K = \alpha - 1$), where $K_c$ characterizes the out of the plane interlayer coupling. This condition is equivalent to the length scale characterizing the vortex pairs, being less than a critical length scale

$$l_0 = \ln(r_0/\xi) = \frac{1}{2}\ln(K/K_c) = \frac{1}{2}\ln(E/E_\perp) \quad (6)$$

where $E$ and $E_\perp$ are the energies to form intralayer and interlayer vortices, respectively. The ratio of the two energy scales is related to the degree of coupling between adjacent superconducting layers. Ideally, $\pi K$ should exhibit a universal jump from 0 to 2 at $T_{BKT}$ for $l = \infty$, but finite $l$ and sample inhomogeneity may smear out a sharp jump. An approximate value of $l_0$ near $T_{BKT}$, can be obtained from the equation given by Hikami and Tsuneto (38), valid for temperatures slightly above $T_{BKT}$,

$$\frac{\Delta K}{\Delta T} = -\frac{4}{\pi^2} l_0 \quad (7)$$

The change of power law exponent with temperature under different pressures are shown in Fig.3(a-d). Pressure dependence of $l_0$ and $K_c$ [See Fig. 3(e, f)] was calculated based on the temperature dependence of power-law exponent $\alpha$ in Fig. 3(b, c). At pressures from 10.0 and 20.4 GPa, the maximum value of $l_0$ (around 5-6) is close to the value of $Tl_2Ba_2CaCu_2O_8$ thin films (with $l_0$=5.3) (39), which was proved to be a 2D superconductor with multilayers. Correspondingly, the weakest interlayer coupling appeared in 10.0 to 20.4 GPa of the A7 phase, with $E_\perp/E'' $ 1, consistent with 2D superconductivity. These results confirm that geometrically three-dimensional systems can behave as two-dimensional systems only if the effect of the coupling between the layers is unimportant. This finding can also be extended to certain cuprates, iron-based superconductors, and nickel-based superconductors with layered crystal structures (6).

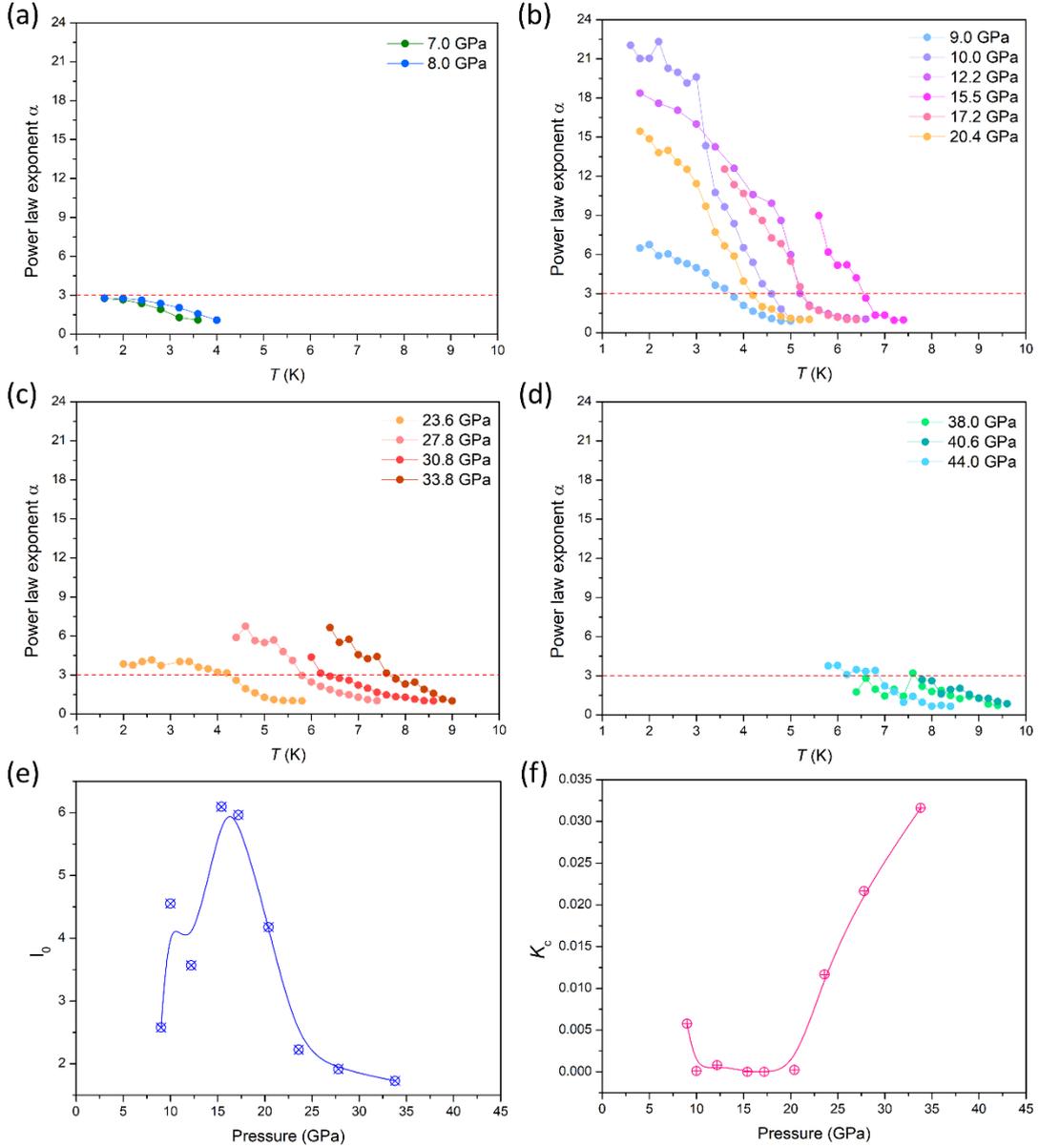

**Fig. 3|** (**a-d**) Power law exponent of the V-I curves for the third run as a function of temperature. Pressure dependence of $l_0$ (**e**) and $K_c$ .(**f**) characterizing the critical length scale and the interlayer coupling in the direction of the out of plane, respectively.

We summarize our experimental results in the pressure-$T_c$ phase diagram in Fig. 4. We observed that the 2D superconducting state and the BKT-transition is above a critical pressure of 9.0 GPa. In addition, high-pressure powder X-ray diffraction measurements at 10 K show that there a clear structural transition from A17 to A7 occurs at ~10.0 GPa. This result implies that the crossover from 3D to 2D superconductivity is related to the pressure-induced phase transition.

When the pressure exceeds 27.8 GPa, $T_c$ increases with a steep drop in the in-plane upper critical field. However, 2D characteristics, such as the BKT transition, were also observed at 27.8 GPa, 30.8 GPa, and 33.8 GPa. When the pressure is more significant than 38.0 GPa, all 2D characteristics disappear. In addition, the mixed phase of A7 and cubic were observed at 30.0 GPa, and 35.0 GPa in our high-pressure powder X-ray diffraction measurements at 10 K, and a pure cubic phase was observed at 40.0 GPa. Therefore, we proposed that the hybrid superconductivity of 2D and 3D is from the presence of mixed crystal structures. The transition from 2D to 3D superconductivity is related to the pressure-induced structural phase transition from A7 to a pure cubic phase.

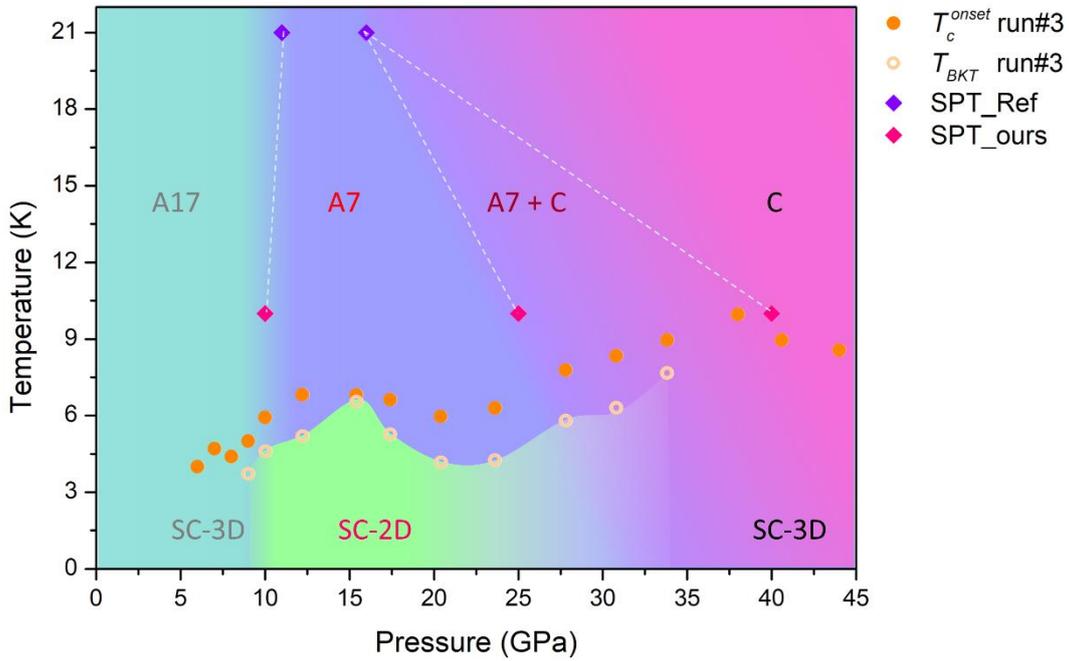

**Fig. 4| The pressure and temperature phase diagram of BP.** SPT (structure phase transition) of Ref. and purple and pink diamonds plot our results. As the pressure increases, the phase changes from A17 to A7 and then to the C phase. This behavior corresponds to the superconducting transitions from 3D to 2D to 3D, respectively.

## 4. Conclusion

We have systematically studied the relationship between the electric transport properties and the structural evolution of BP single crystal under high pressure. The pressure-induced transitions from 3D to 2D to 3D superconducting states were found and the 2D superconducting state can be attributed to the structure of the nanoflakes. This finding has not been reported in any other compressed bulk superconductors and may thus aid in achieving a breakthrough in better understanding layered bulk superconductors and micro-pairing picture. Consequently, we pose some questions stimulated by these results. First, can similar phenomena be observed in other compressed layered bulk superconductors through the same research method as BP?

Next, are there identical phenomena or new quantum phase transitions when the thickness of BP drops to tens of nanometers or a monolayer? Finally, is it possible to uncover an underlying picture to understand the crossover regime from 2D to 3D superconductivity transition through a combination of theoretical and experimental studies?

## Acknowledgements

M. J., Q. W. contributed equally to this work. The authors thank Prof. Lei Shan at Anhui University for the fruitful discussions on 2D superconductivity. The high pressure XRD work was performed at BL10XU of Spring-8, Japan. The work was financially supported by the National Natural Science Foundation of China (No. 12004161, No. 12274193, No. 11921004, No. 12022413, No. 11674331), the National Key R&D Program of China (No.2018YFA0305703, No. 2017YFA0303201), the Guang dong Basic and Applied Basic Research Foundation of 2022A1515010044, the Special Funds for the Cultivation of Guangdong College Students' Scientific and Technological Innovation ("Climbing Program" Special Funds) pdjh2023c20202, the "Strategic Priority Research Program (B)" of the Chinese Academy of Sciences (Grant No. XDB33030100), the Collaborative Innovation Program of Hefei Science Center CAS (No. 2020HSC-CIP002), and the Major Basic Program of Natural Science Foundation of Shandong Province (Grant No. ZR2021ZD01). A portion of this work was supported by the High Magnetic Field Laboratory of Anhui Province, China.